\documentclass[a4paper]{PoS}

\usepackage{amsmath} 
\usepackage{epsfig}
\usepackage{subfig} 

\usepackage{ulem} 
\usepackage[babel=true]{csquotes}    

\newcommand{\beq}{\begin{eqnarray}}
\newcommand{\eeq}{\end{eqnarray}}
\newcommand{\be}{\begin{eqnarray*}}
\newcommand{\ee}{\end{eqnarray*}}

\newcommand{\ie}{{\it i.e.}}

\newcommand{\etal}{{\it et al.}}

\newcommand{\cf}[1]{{Fig.~\ref{#1}}}

\def\lsim{\raise0.3ex\hbox{$<$\kern-0.75em\raise-1.1ex\hbox{$\sim$}}}
\def\gsim{\raise0.3ex\hbox{$>$\kern-0.75em\raise-1.1ex\hbox{$\sim$}}}

\def\beq     {\begin{equation}}
\def\eeq     {\end{equation}}

\title{{$\psi(2S)\hspace*{-1.535cm}\psi(2S)$} production in proton-proton collisions at RHIC, Tevatron and LHC energies}

\ShortTitle{$\psi(2S)$ production in proton-proton collisions at RHIC, Tevatron and LHC energies}

\author{\speaker{J.P.~Lansberg}\\
        IPNO, Universit\'e Paris-Sud, CNRS/IN2P3, F-91406, Orsay, France\\
        E-mail: \email{lansberg@in2p3.fr}

        }

\abstract{
We briefly review the existing $\psi(2S)$ data taken at RHIC, the Tevatron and the LHC. We systematically compare them with
colour-singlet-model predictions as a function of the center-of-mass energy, of the quarkonium rapidity and of the quarkonium 
transverse momentum. The overall agreement is good except for large transverse momenta. This points at the existence 
of large NNLO corrections or points at colour-octet dominance.
}

\FullConference{36th International Conference on High Energy Physics\\
		 4-11 July 2012\\
		 Melbourne, Australia}

\begin{document}


\section{Introduction}

The physics of quarkonium production in high-energy hadron collisions has triggered much investigation and debate
since the preliminary data of prompt $\psi(2S)$ production at the Tevatron were released by 
the CDF collaboration in 1994~\cite{CDF:1994aa}.
It uncovered an obvious discrepancy between the measured $P_T$ spectrum  and the one predicted by
the simple application of pQCD, dubbed the colour-singlet model (CSM): the spectrum of the data was harder with a clear excess
of the $\psi(2S)$ at large $P_T$. This observation was confirmed in the final publication in 1997~\cite{Abe:1997jz} and similar 
observations were also made for the $J/\psi$ --once the feed-down from the $\chi_c$ 
could be subtracted~\cite{Abe:1997yz}-- and for the $\Upsilon$ family~\cite{Abe:1995an} --albeit to a less extent.

Until not loo long ago, the most popular explanation of these discrepancies was provided by 
an enhancement of the yield due to non-perturbative transitions between the colour-octet state
produced at short distances and the colour-singlet mesonic states (for reviews see~\cite{Lansberg:2006dh}). However, it is clear nowadays that one cannot neglect the 
$\alpha^4_S$ and $\alpha^5_S$ corrections to the CSM~\cite{CSM_hadron} 
if one wants to confront the prediction of the CSM with the data for the $P_T$ spectrum of $J/\psi$ and $\Upsilon$ produced in
high-energy hadron  collisions~\cite{Campbell:2007ws,Artoisenet:2007xi,Gong:2008sn,Artoisenet:2008fc,Lansberg:2008gk}.
In fact, this indicates that a factorised description of high-$P_T$ quarkonia beyond leading power~\cite{Kang:2011mg}
may be more suitable than NRQCD~\cite{Bodwin:1994jh}.

QCD corrections also have a dramatic impact on the predictions of polarisation observables. At LO,  the 
yield of inclusive $J/\psi$ and $\Upsilon$  is transverse --in the helicity frame--, whereas
it "becomes" longitudinal at NLO at large $P_T$~\cite{Gong:2008sn,Artoisenet:2008fc}. This is also true for instance
when they are produced in association with a photon~\cite{Li:2008ym,Lansberg:2009db}. The sole known case
where the quarkonium polarisation is not altered by NLO corrections is when they are produced with a
$Z$-boson~\cite{Gong:2012ah}.

On the other hand, in recent works~\cite{Brodsky:2009cf,Lansberg:2010cn}, we have shown that the CSM  alone is 
sufficient to account for the magnitude of the $P_T$-integrated cross section and its dependence in rapidity, 
$d\sigma/dy$, at RHIC, Tevatron and LHC energies. In other words, there is no need for additional contributions at
low $P_T$ and the energy dependence is well reproduced. 

We consider here  the $\sqrt{s}$, $P_T$ and $y$ dependences 
of the $\psi(2S)$ yield at collider energies. We present various comparisons between the existing data and
the yield at LO, NLO and sometimes including some dominant contributions at $\alpha_S^5$ (NNLO$^\star$) when possible.
It should be stressed that the NLO code for the CSM at high energies and low $P_T$ is not stable likely owing to large contributions
of the loop corrections which can become negative at low $P_T$. NLO and NNLO$^\star$ are thus only computed
for sufficiently large $P_T$.

After having briefly described how the $\psi(2S)$ cross sections are evaluated in the CSM, we compare them with
the existing data as a function of the cms energy, of the quarkonium rapidity and of its transverse momentum.
As we shall see, the overall agreement is good except for large transverse momenta. 

\section{Cross-section evaluation in the CSM}

Details about the evaluation of quarkonium production cross section at LO in the CSM can be found in~\cite{CSM_hadron}. As regards
the evaluation of the yield at NLO accuracy, we have used the code of Campbell, Maltoni and Tramontano~\cite{Campbell:2007ws}.
In the CSM, the cross section to produce a $\psi(2S)$ is obtained along the same lines as for the $J/\psi$ ground state
with the mere change of the value of the wave function at the origin $R(0)$, which is meant to account for all the
non-perturbative and relativistic effects. For the $J/\psi$, $|R(0)|^2$ can be taken as 1.01 GeV$^3$; here, we took
0.67 GeV$^3$ for the $\psi(2S)$~\cite{Brodsky:2009cf}.

The expected impact of NNLO QCD corrections for increasing $P_T$ is investigated by evaluating 
what we call the NNLO$^\star$ yield. A complete discussion can be found in~\cite{Artoisenet:2008fc,Lansberg:2008gk}.
In a few words, we anticipate that the NNLO$^\star$ yield encompasses the kinematically-enhanced topologies which open up
at $\alpha^5_S$  with a $P_T^{-4}$ fall off of $d\sigma/dP_T^2$. Let us emphasise that we do not foresee significant 
modifications of the $P_T$ dependence  at N$^3$LO and further. At NNLO, the  $P_T^{-4}$ fall-off of the new NNLO 
topologies is the slowest possible. Above $\alpha_S^5$, the common wisdom about the decreasing 
impact of further QCD corrections would then hold. One expects a $K$ factor multiplying the NNLO yield to be independent of $P_T$ 
and to be of the order of unity.
It would be worrisome to find out a further enhancement  between the NNLO and N$^3$LO results by an order of magnitude.  
As in~\cite{Artoisenet:2008fc,Lansberg:2008gk}, it  is evaluated thanks to a slightly tuned
version of the automated code {\small MADONIA}~\cite{Artoisenet:2007qm}.

The uncertainty bands at LO and NLO are obtained from the {\it combined}
 variations of the charm-quark mass ($m_c=1.5\pm 0.1$ GeV), the
factorisation $\mu_F$ and the renormalisation $\mu_R$ scales
 chosen in the couples $((0.75,0.75);(1,1);(1,2);(2,1);(2,2))\times m_T$ with $m^2_T=4m_Q^2+P_T^2$.
The band for the NNLO$^\star$ is obtained using a combined variation of $m_c$, $0.5 m_T <\mu_R=\mu_F< 2 m_T$
and $2.25  < s_{ij}^{\rm min}<   9.00$ GeV$^2$. We have used 
the LO set {\small CTEQ6\_L} and the NLO set {\small CTEQ6\_M}~\cite{Pumplin:2002vw} and have taken $|R_{\psi(2S)}(0)|^2=0.67$ GeV$^3$ and 
Br$(\psi(2S) \to \ell^+\ell^-)=0.0075$.

\section{Result and comparison with existing data}
\label{sec:partonic-process}
\subsection{$P_T$-integrated yields}
In \cf{fig:x-sect-vs-s}, we compare the $\sqrt{s}$ dependence of  $d\sigma/dy|_{y=0}$ obtained at LO in the CSM
and that obtained from the PHENIX data ($\sqrt{s}=200$~GeV)~\cite{Adare:2011vq}, the CDF data\footnote{To be precise,
the CDF data only covers the region $P_T> 2$~GeV, for which $\sigma(|y|< 0.6,P_T> 2\hbox{ GeV})\times \hbox{Br}=2.6 \pm 0.1$~nb.
We have assumed the same $P_T$ dependence as the inclusive $J/\psi$ for $P_T> 2$~GeV~\cite{Acosta:2004yw} and we have obtained
$\sigma(|y|< 0.6,P_T < 2\hbox{ GeV})\times \hbox{Br}=2.1 \pm 0.1$~nb. On the way, it is instructive to
keep in mind that 45\% of the $J/\psi$ yield lies below $P_T=2$~GeV.}   ($\sqrt{s}=1960$~GeV)~\cite{Aaltonen:2009dm} and rescaled LHCb data 
($\sqrt{s}=7000$ GeV)~\cite{Aaij:2012ag}.
As one has obtained for the $J/\psi$ and the $\Upsilon(nS)$~\cite{Lansberg:2010cn},
 one does not observe any surplus w.r.t. to the CSM predictions. In fact, at high energy, the 
LO yield tends to be above the experimental data. 

\begin{figure}[htb!]
\begin{center}
\includegraphics[width=.4\linewidth,draft=false]{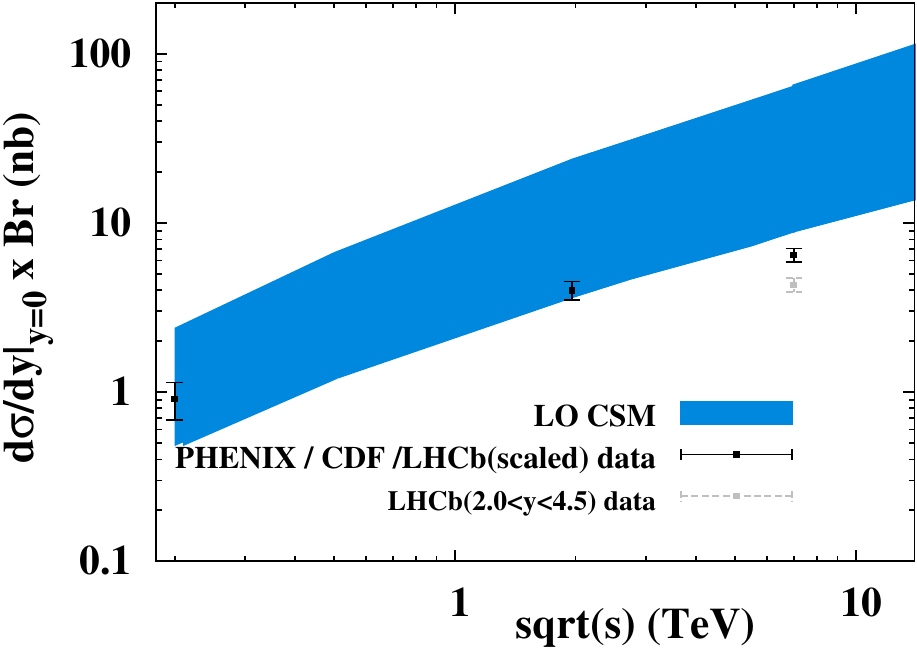}
\vspace*{-.5cm}
\end{center}
\caption{Data vs LO CSM for the $P_T$-integrated $\psi(2S)$ production cross section
at $y=0$ as a function of the cms energy. The forward LHCb data are rescaled assuming the same $y$ dependence
as the LO CSM, \ie~using a factor 1.5 for $d\sigma/dy|_{y=0}/\langle d\sigma/dy\rangle|_{2.0<y<4.5}$.
Data are from~\cite{Adare:2011vq,Aaltonen:2009dm,Aaij:2012ag}.}
\label{fig:x-sect-vs-s}
\end{figure}

\begin{figure}[htb!]
\begin{center}
%
\subfloat[][RHIC]
{%
\label{fig:x-sect-vs-y-a}
\includegraphics[width=.45\linewidth]{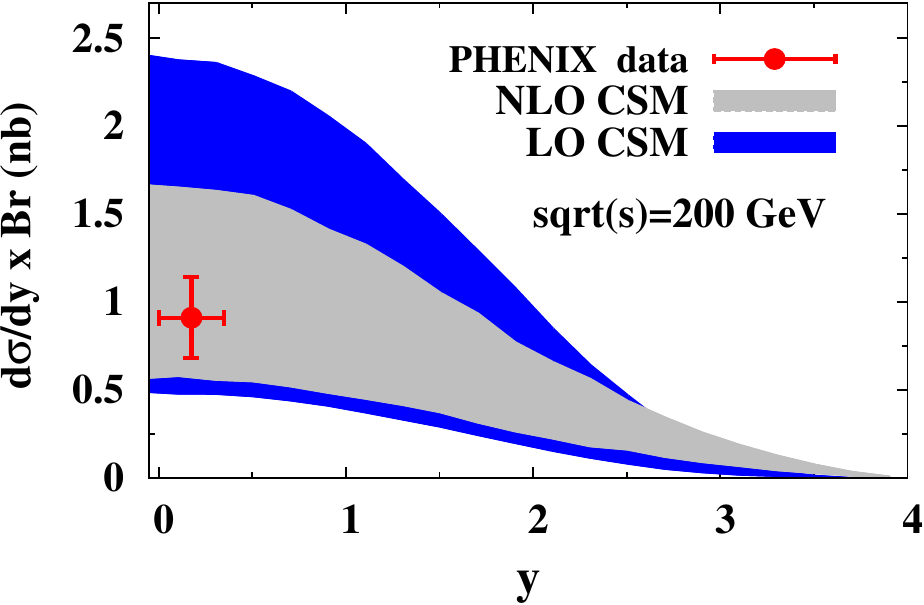}}
\subfloat[][LHCb]
{%
\label{fig:x-sect-vs-y-b}
\includegraphics[width=.45\linewidth]{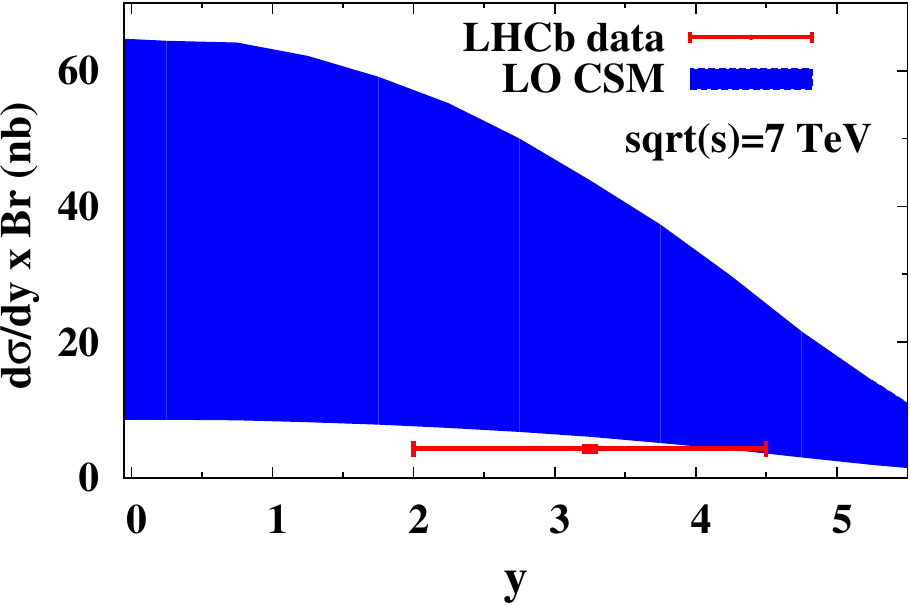}}
\vspace*{-.5cm}
\end{center}
\caption{Data vs  (N)LO CSM for the $P_T$-integrated direct-quarkonium production cross section
as a function of the $\psi(2S)$ rapidity.
Data are from~\cite{Adare:2011vq,Aaij:2012ag}.
}
\end{figure}

As mentioned above, we are unfortunately not able to give, for the time being, corresponding predictions at NLO since 
these are not well behaved at high energies. This is maybe due to the contributions
of the loop corrections which change sign for $P_T$ of the order of the quark masses. This calls for
the resummation of initial state radiations (ISR). Yet, at RHIC energies, we have observed in~\cite{Brodsky:2009cf}
that the NLO yield lies in the lower range of the LO uncertainties (see \cf{fig:x-sect-vs-y-a}). It is therefore sound to
expect that the slight overestimation of the LO w.r.t. the present data would be reduced at NLO once ISR can 
be resummed and stable results can be presented. \cf{fig:x-sect-vs-y-a} and \cf{fig:x-sect-vs-y-b} shows the
rapidity dependence at Tevatron and LHC energies. Note that, for now, there is no measurement of the $\psi(2S)$ yield
for low enough $P_T$ to derive a $P_T$-integrated cross section for $y \simeq 0$.

\subsection{$P_T$-differential yields}

Now, we move onto the discussion of the $P_T$ dependence of the cross section. As we discussed in the introduction, 
it is clear that the LO CSM cannot be sufficient since the leading (and even sub-leading) $P_T$ topologies are missing.
Except for low $P_T$, where the results are not stable, the NLO yield can nowadays be computed easily and compared to the data.
It is shown at RHIC and Tevatron energies in \cf{fig:x-sect-vs-pt-a} and \cf{fig:x-sect-vs-pt-b}; one clearly sees
a different $P_T$ dependence w.r.t. the LO yield. It is clearly harder and the discrepancy with the experimental
data is reduced, altough it is still significant when $P_T$ gets large. 

\begin{figure}[htb!]
\begin{center}
\subfloat[][RHIC]
{%
\label{fig:x-sect-vs-pt-a}
\includegraphics[width=.4\linewidth]{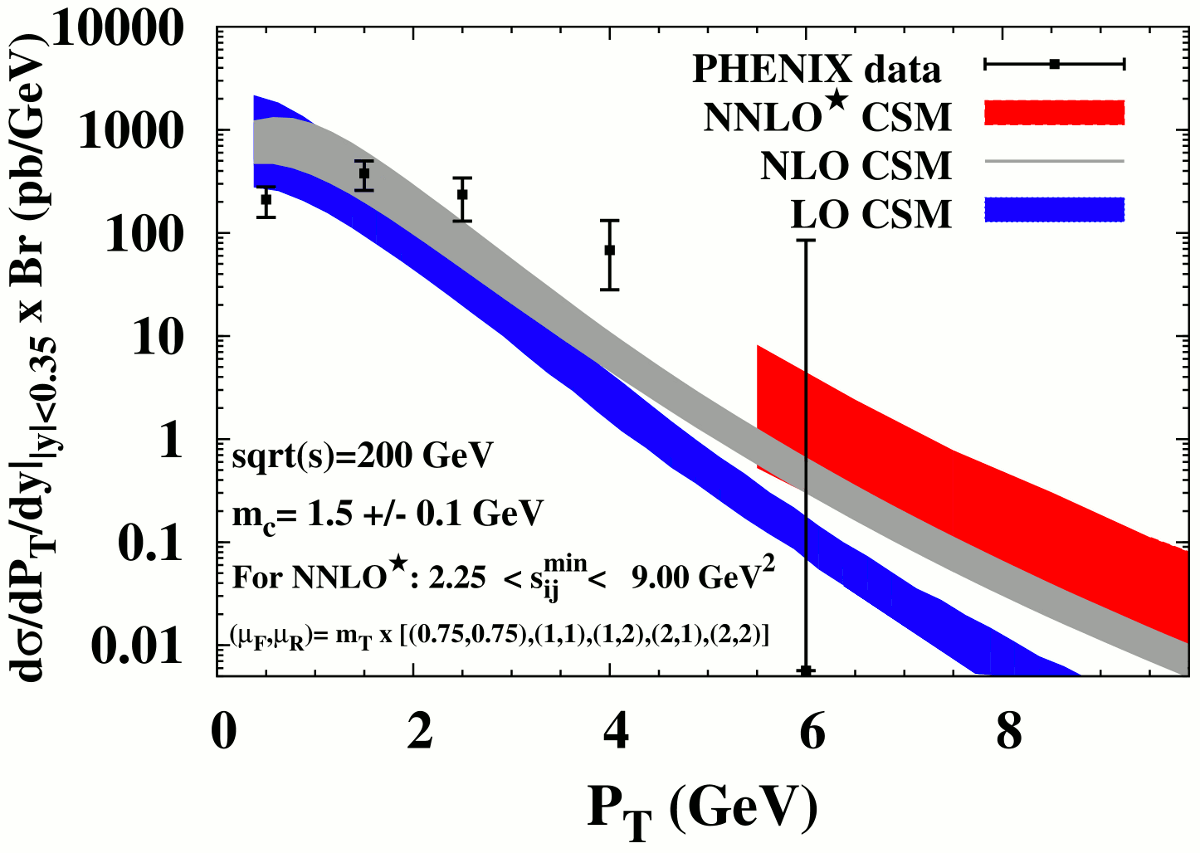}}\hspace*{-.1cm}
\subfloat[][Tevatron]
{%
\label{fig:x-sect-vs-pt-b}
\includegraphics[width=.41\linewidth,draft=false]{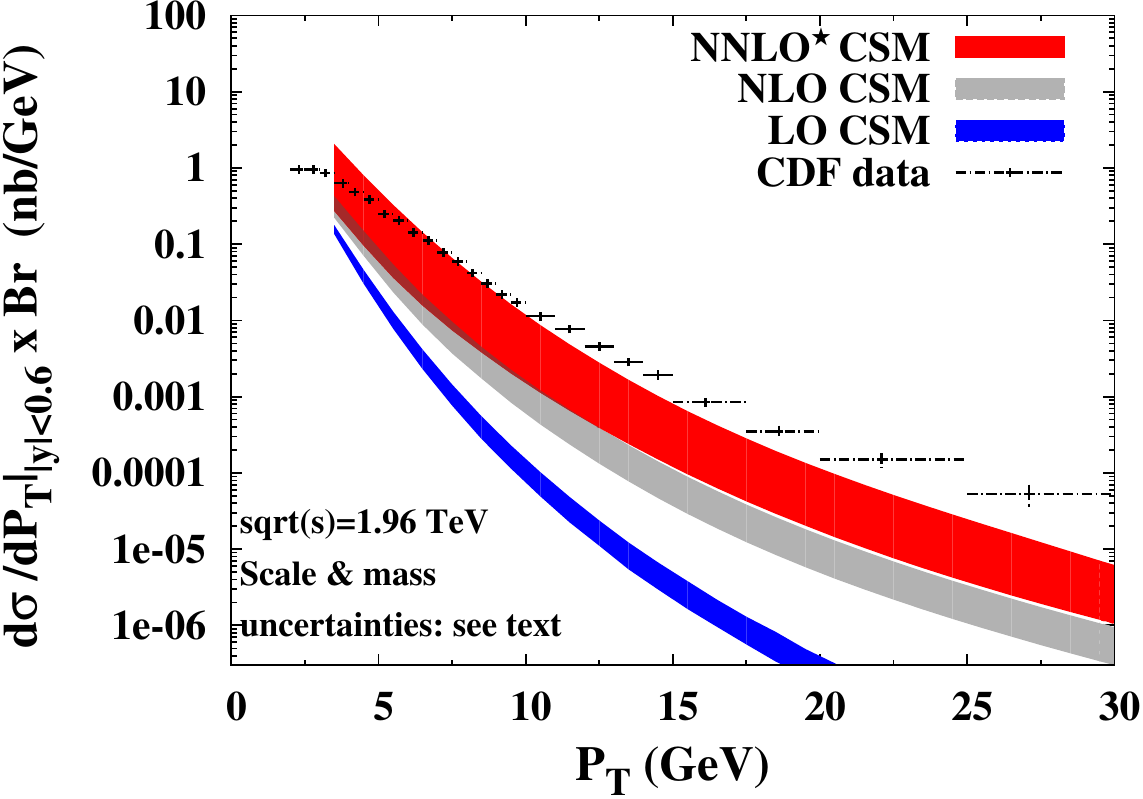}}\hspace*{-.1cm}\\
\subfloat[][LHC CMS]
{%
\label{fig:x-sect-vs-pt-c}
\includegraphics[width=.4\linewidth,draft=false]{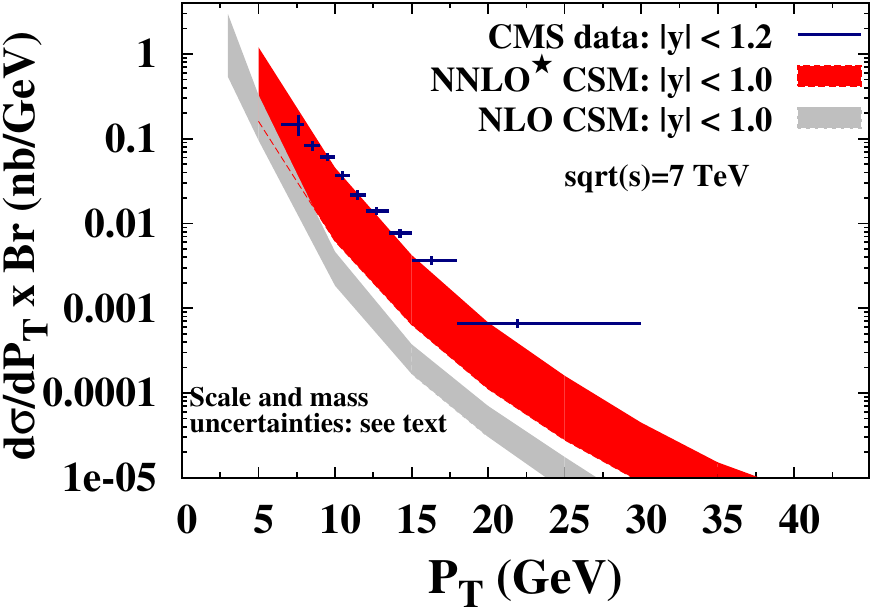}}\hspace*{-.1cm}
\subfloat[][LHC LHCb]
{%
\label{fig:x-sect-vs-pt-d}
\includegraphics[width=.43\linewidth]{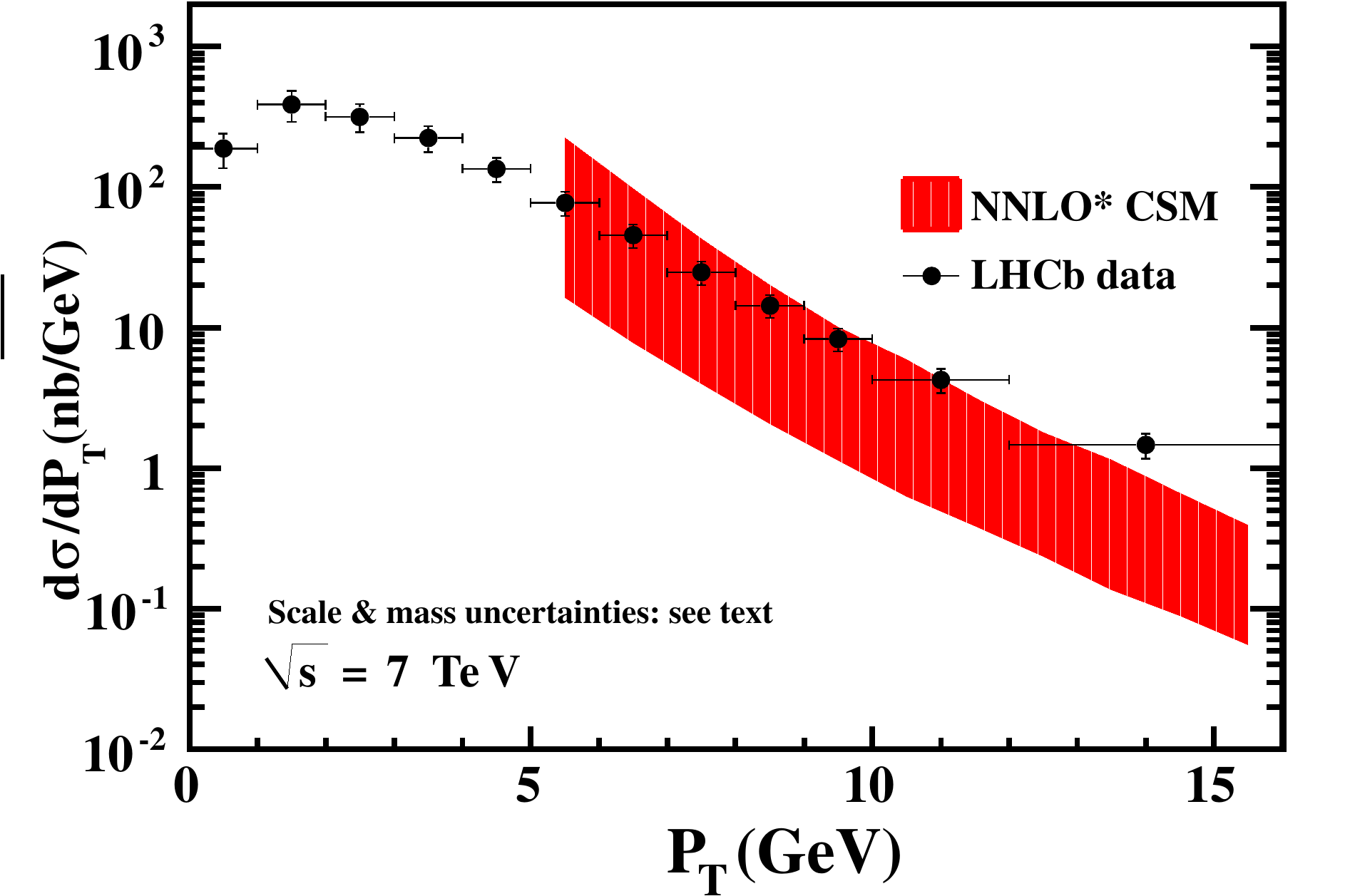}}\hspace*{-.1cm}
\vspace*{-.5cm}
\end{center}
\caption{$\psi(2S)$ data vs CSM predictions at various orders for the   $P_T$ dependence of the differential cross section
at RHIC, Tevatron and LHC energies. Data are from~\cite{Adare:2011vq,Aaltonen:2009dm,Aaij:2012ag,Chatrchyan:2011kc}.}
\end{figure}

Indeed, at large $P_T$, one expects the $P_T^2$ kinematical enhancement of some NNLO topologies to take over the extra
$\alpha_s$ suppression w.r.t. the NLO. This is why the NNLO$^\star$ bands in \cf{fig:x-sect-vs-pt-a}-\ref{fig:x-sect-vs-pt-d} 
show a $P_T$ spectrum which is even harder and much closer to the 
data. If the CSM at NNLO is indeed the physically relevant production mechanism at large $P_T$, 
namely that the charmonia are produced at large $P_T$ along
with two hard partons (in a sense, two jets), the corresponding theoretical predictions would be involving. Even
with a full NNLO, one would have to face large theoretical uncertainties due to the factorisation scale through five powers
of $\alpha_s$. In the coming years, it is not clear that one could make precise and definitive comparisons between
data and theory as far as the $P_T$ dependence of the yield is concerned. It may thus be more fruitful to also analyse
new observables such as associated production.

For the time being, the most one can tell from the data-theory comparison in the framework of the CSM and the approximate
NNLO$^\star$ is that below, say $P_T=15$ GeV, the upper uncertainty band of the NNLO$^\star$ agrees with the data -- the same 
observation also holds for the $J/\psi$~\cite{Lansberg:2008gk,Lansberg:2010vq,Lansberg:2011hi} while, for the $\Upsilon(nS)$, 
the agreement is clearly much better~\cite{Artoisenet:2008fc,Lansberg:2012ta}. Whether this
has some physical meaning is a question that would only be answered once one has a full NNLO computation and/or once
the LHC experimental collaborations release polarisation measurement for prompt $\psi(2S)$ which would be 
precise enough to rule out the predictions from either the CSM or from the CO dominance.

\section{Conclusion}

We have compared the existing data of $\psi(2S)$ production at RHIC, the Tevatron and the LHC 
with predictions from the CSM for the $P_T$, $\sqrt{s}$, $y$ dependences of the yield.
The $\sqrt{s}$ and $y$ dependences are well reproduced at LO and NLO, where it is available.
It is therefore worth re-investigating~\cite{Lansberg:2012kf,Brodsky:2012vg,Diakonov:2012vb} 
the possibilities to constrain gluon PDFs with low 
$P_T$ quarkonium data.

As regards the $P_T$ differential cross section, the upper bound of the NNLO$^\star$ CSM 
predictions of the $P_T$ differential cross section 
is very close to the experimental data, as previously   found for $\Upsilon$~\cite{Artoisenet:2008fc,Lansberg:2012ta}  and for
 $J/\psi$~\cite{Lansberg:2008gk,Lansberg:2010vq,Lansberg:2011hi}. 
However, the NNLO$^\star$ evaluation is not a complete NNLO calculation. 
It is affected by logarithms of an infrared cut-off and its effect might not vanish 
with increasing $P_T$ as quickly as one has anticipated. A full NNLO evaluation of the cross section in the CSM
is therefore awaited for.

\subsubsection*{Acknowledgments}

I thank P. Artoisenet, S.J. Brodsky, J. Campbell, F. Maltoni and F. Tramontano for
our fruitful collaborations on some of the topics presented here. I also thank J. He for useful discussions and
for providing me with some data.



\end{document}